\documentclass[aps,pre,showpacs,twocolumn]{revtex4}

\usepackage{epsfig,amsmath,amssymb,bm,epsf,graphics}
\usepackage{mathrsfs,float}
\begin{document}

\title{Effects of Disorder in Location and Size of Fence Barriers on \\ 
Molecular motion in Cell Membranes}

\author{Z. Kalay}
\affiliation{Consortium of the Americas for Interdisciplinary Science  
and
Department of Physics and Astronomy, University of New Mexico,  
Albuquerque,
New Mexico 87131, USA}
%\date{\today}

\author{P. E. Parris}
\affiliation{Consortium of the Americas for Interdisciplinary Science  
and
Department of Physics and Astronomy, University of New Mexico,  
Albuquerque,
New Mexico 87131, USA}

\author{V. M. Kenkre}
\affiliation{Consortium of the Americas for Interdisciplinary Science  
and
Department of Physics and Astronomy, University of New Mexico,  
Albuquerque,
New Mexico 87131, USA}

\begin{abstract}
The effect of disorder in the energetic heights and in the physical  
locations of fence barriers encountered by transmembrane molecules such  
as proteins and lipids in their motion in cell membranes is studied  
theoretically. The investigation takes as its starting point a recent  
analysis 
%(Kenkre, Giuggioli, Kalay, Phys. Rev. E, submitted for publication) 
of a periodic system with constant  
distances between barriers and constant values of barrier heights, and  
employs effective medium theory to treat the disorder. The calculations  
make possible, in principle, the extraction of confinement  
parameters such as mean compartment sizes and mean intercompartmental transition rates from experimentally reported published  
observations. The analysis should be helpful both as an unusual  
application of effective medium theory and as an investigation of observed molecular movements in cell  
membranes.
\end{abstract}

\maketitle

\section{Introduction}
\label{intro}
The biophysics of cell membranes is an active field of current  
research, issues of interest being cell shaping and movement  
\cite{mcmahongallop}, cell division \cite{celldiv}, signal transduction  
\cite{kraussbook}, and molecule trafficking \cite
{maxfieldtabas}. Observations of the lateral movement of molecules on  
the surface of the
cell  
\cite{edidin91,edidin94,sako98,tomishige,simson98,capps04,kusumireview}  
  have given rise to the idea that the moving (transmembrane) molecules  
are confined within certain regions of
the cell membrane. One possible source of this confinement has been suggested
\cite{kusumireview} as being collisions of membrane molecules  
protruding into the cytoplasm with the cytoskeleton \cite{howardbook}.  
The model views the
molecules  as moving freely, their motion being hampered as they  
traverse
adjacent compartments. As the actin filament that forms the compartment
boundary dissociates due to thermal fluctuations, the  moving molecules is envisaged as overcoming the barrier potential and  
hopping
to the adjacent compartment. 

To our knowledge, two theoretical attempts have been made in the  
literature to analyze such barrier-hindered motion of transmembrane  
molecules. In both of them the molecule is looked upon as a random  
walker moving (in a 1-D system for simplicity) with periodically  
arranged semipermeable barriers. The more recent of the two attempts  
borrows the spirit of  the earlier one \cite{powlesetal}   but avoids some of its  
shortcomings. The shortcomings include the appearance of an unfortunately  
unwieldy infinite series of terms which is difficult to handle, and the  
unavailability of explicit usable expressions for the mean square  
displacement, which is the quantity of direct comparison to the  
experiment \cite{kusumireview}. It is that second (more recent)  
theoretical attempt  \cite{KGK} that we take as our starting point here  
and calculate by substantial modifications of that analysis the  
consequences of disorder on the effective diffusion constant and the molecular mean square displacement.

In the model analyzed in ref. \cite{KGK}, we represent the molecule as a random walker in a 1-D infinite chain of sites. The molecule, whose probability of occupation of the $m$-th site of the chain at time $t$ is $
P_{m}(t)$, hops via nearest neighbor transfer rates. The rate is $F$ within a compartment and has a lower value $f$ at the interface of compartments where there is a barrier hindering the molecular motion. There are  $H+1$
sites, equivalently $H$ nearest neighbor bonds,  within a compartment; for simplicity, $H$ is taken to be even, with the site $0$ at the center of one of the
compartments. Specifically, the Master equation
\begin{eqnarray}
\frac{dP_{m}}{dt}=F\left[ P_{m+1}+P_{m-1}-2P_{m}\right] \nonumber \\ -(F-f) \sum_{r}^{\prime}\left[ P_{r+1}-P_{r}\right] \left( \delta _{m,r}-\delta
_{m,r+1}\right)  \label{1stME}
\end{eqnarray}
describes that ordered system. The primed summation goes over sites $r=H/2+(H+1)l$ which lie to the left of each barrier, $l$ taking all integer values.

Explicit expressions have been provided in ref. \cite{KGK} for the time dependence of the mean square displacement of the molecule and for the  effective diffusion constant, the compartment size $H$ and the transfer rates $f$ and $F$ being reflected transparently in these calculated quantities. These are based on Eqs. (\ref{e2}) to appear in Section 2 below. The major element missing from that \emph{periodic barrier} theory is  
the realistic effect of the compartment sizes and the barrier heights  
being not equal throughout the system, in other words of $H$ and $f$  
being variable, i.e., disordered, quantities. Our interest in the present paper is to  
treat this important disorder effect. The tool we employ is  effective medium theory: a disordered system is  
replaced by an ordered system properly structured to represent the  
original system. There is an extensive literature on the general  
subject of effective medium theory\cite%
{kirkpatrick,odagakilax,hauskehr,parris,kenkre}, and there is a long
history of its application to problems involving normal and anomalous  
transport
in disordered lattices. Recently, the approach has been applied with  
success to the study of transport in more complex environments,  
including
random graphs \cite{zwanzig} and small world networks \cite{parriskenkre,cpk, pck}. Some of those developments are naturally applicable to the present problem of molecular motion in cell membranes as we shall see below.

The motivation to extend our previous theory \cite{KGK} to treat disordered $f$ and $H$ stems from the experimentally known fact that the barrier locations for the transmembrane molecular motion in the cellular membrane occur at positions that are by no means regular; the consequent variations are substantial within a system, about one order of magnitude, the compartment sizes  sometimes being quoted as lying between $30 \, nm$ and $240 \, nm$ \cite{kusumireview}. The precise situation in which the moving molecules find themselves at the barriers also varies, the result being a variation in the effective transfer rate.

We construct our effective medium theory considerations for the present  
problem in three parts.  First, in Section 2, we take the compartment  
sizes to be all equal as in ref. \cite{powlesetal} or \cite{KGK} but  
allow the barrier heights, consequently the inter-compartment rates of  
molecular motion $f$, to be taken from each of several specific distribution  
functions with given mean, variance, and nature. In Section 3, we first take  
the inter-compartment rates to be  constant throughout, but  
allow the compartment sizes $H$ to vary, and then allow both  
$H$ and $f$ to be random variables. Fig. 1 illustrates the three respective cases as (a), (b) and (c). 
\begin{figure}[H]
%\centering
\includegraphics[width = 8.0cm]{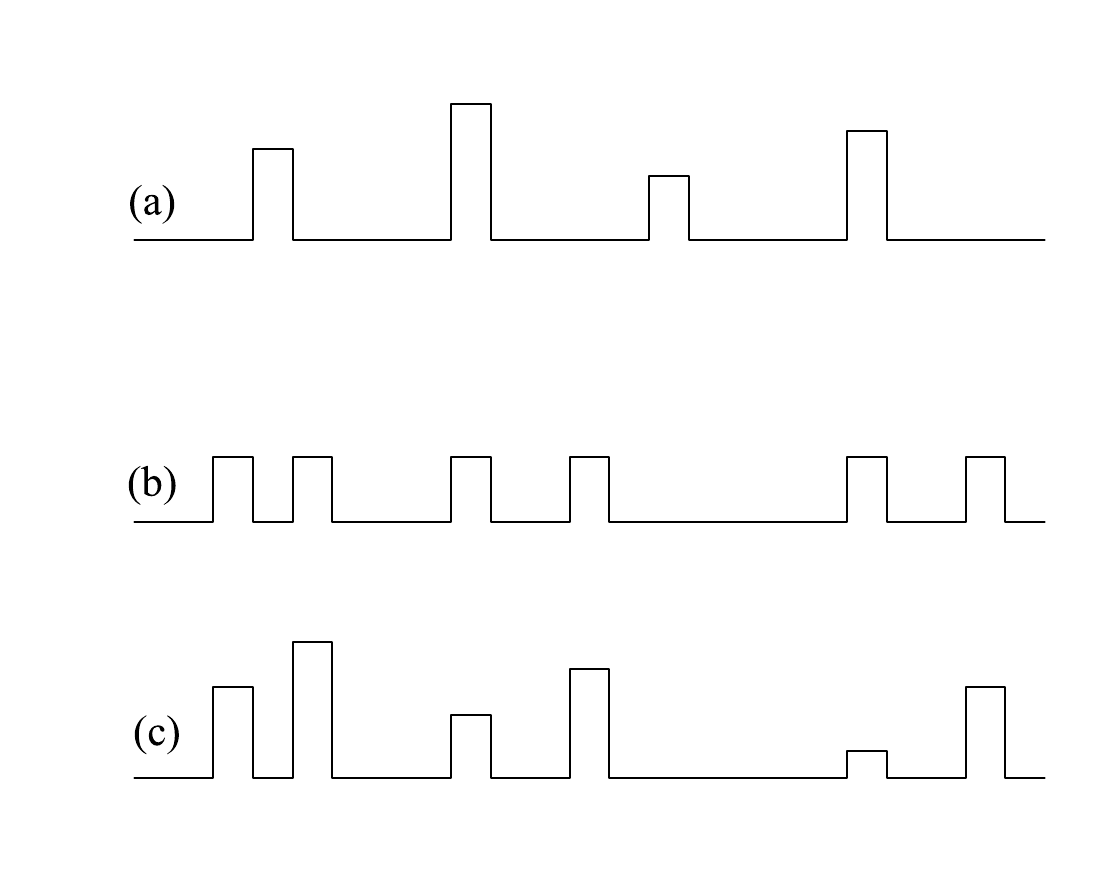}
\caption{Schematic illustration of the different types of barrier disorder considered. In (a), the barrierÕs are periodically-spaced, but have random energetic heights. In (b), the barrier heights are uniform, but the distances between barriers is random. In (c), both the barrier spacing and the barrier heights are independent random variables.}
\label{DisorderTypes}
\end{figure}
In each case we calculate physical observables typified by the effective diffusion constant. Whereas the $f$ distributions we  
consider in Section 2 are arbitrary, the $H$-distributions we consider  
in Section are not arbitrary but determined by the specific $f$  
distributions we take to generate them. This allows us to make direct  
use of the analysis (in particular, the form of the  transport propagators)  developed in  
ref. \cite{KGK}. Fully arbitrary distributions for compartment size will be  
studied in a future publication. A graphical comparison of the predictions of our effective medium theory with the results of a numerical solution of the disordered problem, and  concluding remarks are presented in  
Section 4. A brief analysis of the memory function that emerges from the effective medium theory is also given in the discussion in Section 2.

\section{Disorder in Barrier Heights}
\label{disorderheights}
The system we investigate in the present paper is  formally described by Eq. (\ref{1stME}) as in ref. \cite{KGK} but with the understanding that $f$ and the primed locations $r$ are disordered quantities. In the analysis of ref. \cite{KGK}, the solution of the ordered Eq. (\ref{1stME}) for arbitrary initial probabilities $P_n(0)$ is expressed as
$$P_m(t)=\sum_n \chi_{m,n}(t)P_n(0)$$
where the transport propagator (which `propagates' the solution from site $n$ to site $m$) is given in the Laplace domain ($\epsilon$ being the Laplace variable and tildes denoting Laplace transforms) by
\begin{align}
\label{e2}
\widetilde{\chi}_{m,n}&=\widetilde{\Psi}_{m-n}-\frac{F-f}{1+(F-f)\widetilde{\mu}}\\ &\times \sum_{r}^{\prime}(\widetilde{\Psi}_{r-n+1}-\widetilde{\Psi}_{r-n})(\widetilde{\Psi}_{m-r}-\widetilde{\Psi}_{m-r-1}). \nonumber
\end{align}
This is a slightly rewritten, but completely equivalent, version of Eq. (2) of ref. \cite{KGK}.

Here, the first term on the right hand side is the Laplace transform of the propagator $\Psi_{m-n}(t)$ of the system without barriers ($F=f$), and is characterized by a single index as a result of complete translational invariance at the site level. The second term describes the effect of the barriers between compartments. It is proportional to the difference $F-f$ and is characterized by a property of the barrierless system, viz., products of the propagator differences in the Laplace domain, summed over barrier locations, and is also characterized by $\mu(t)$, an appropriate summed combination of $\Psi(t) $'s (see ref. \cite{KGK}), whose Laplace transform is given by
$$\widetilde{\mu }(\epsilon
)=\frac{1}{F}\left[ \frac{\tanh \left( \xi /2\right) }{\tanh \left( \xi\left( H+1\right) /2\right) }-1\right] ,
$$
with $\xi=2\sinh^{-1}(\sqrt{\epsilon / 4F})$. In obtaining this result, one uses the known form of the Laplace transforms of the propagators for the system without barriers,
\begin{equation}
\widetilde{\Psi}_l=\frac{e^{-\xi|l|}}{2F\sinh \xi}. 
\label{hyperbolic}
\end{equation}

\subsection{General Development}
Our goal in this Section is to analyze the generalization of the system represented by Eq. (\ref{e2}) when the intercompartmental transition rates $f$ vary in magnitude throughout the chain and are picked from a distribution function $\rho(f)$. The equation obeyed by the probabilities of occupation is
\begin{eqnarray}
\frac{dP_{m}}{dt}=F\left[ P_{m+1}+P_{m-1}-2P_{m}\right] \nonumber \\ - \sum_{r}^{\prime}(F-f_r)\left[ P_{r+1}-P_{r}\right] \left( \delta _{m,r}-\delta
_{m,r+1}\right)  \label{tougheqn}
\end{eqnarray}
An exact analytic solution of Eq. (\ref{tougheqn}) is practically impossible for large systems because the $f_r$'s vary in a disordered fashion. Therefore, in the spirit of effective medium theory \cite{kirkpatrick,odagakilax,hauskehr,parris,kenkre}, we replace the actual 
disordered system with its many $f$'s by an effective medium system characterized by a single quantity (memory function) $\mathcal{F}(t)$  which is time-dependent and to be determined from the distribution $\rho(f)$. The effective medium system is identical to the periodic system represented by Eq. (\ref{e2}) except that $f$ is replaced by $\mathcal{\widetilde{F}(\epsilon)}$.  This replacement means that the occupation probabilities in the effective medium system obey
\begin{gather}
\frac{dP_{m}(t)}{dt}=F\left[P_{m+1}(t)+P_{m-1}(t)-2P_{m}(t)\right] \nonumber \\ -\int_{0}^{t}dt^{\prime}\left[F\delta(t-t^{\prime})-\mathcal{F}(t-t^{\prime})\right] \nonumber \\ \times\sum_{r}^{\prime}\left[ P_{r+1}(t^{\prime})-P_{r}(t^{\prime}) \right](\delta_{m,r}-\delta_{m,r+1}) 
\label{teff}
\end{gather}
Equation (\ref{teff}) differs from the ordered counterpart (\ref{1stME}) in just one aspect: the appearance of the effective memory $\mathcal{F}(t)$ in place of the constant $f$. This memory, and the consequent convolution form for the equation, is an essential and well-known feature of the effective medium treatment. For long-time considerations, one replaces
$\mathcal{F}(t)$ by its Markoffian
 approximation $\delta(t)\int_0^\infty \mathcal{F}(t^{\prime})dt^{\prime}$. Then the formal identity to Eq. (\ref{1stME}) is exact.
 
To determine $\mathcal{F}$, we follow the effective medium prescription \cite{kirkpatrick,odagakilax,hauskehr,parris,kenkre} of considering one defect in the otherwise periodic system (\ref{teff}) formed by replacing $\mathcal{F}$ by an $f$ drawn from its probability distribution $\rho(f)$, solving the defect problem exactly in the Laplace domain, averaging the solution over the $f$'s in the distribution, i.e., carrying out an ensemble average of the solutions, and then requiring that the ensemble-averaged solution is equal to the solution of the system without the defect. 

The propagator for the effective medium system (\ref{teff}) is, in the Laplace domain,
\begin{gather}
\widetilde{\chi}_{m,n}=\widetilde{\Psi}_{m-n}-\frac{F-\widetilde{\mathcal{F}}}{1+(F-\widetilde{\mathcal{F}})\widetilde{\mu}} \nonumber \\ \times\sum_{r}^{\prime}(\widetilde{\Psi}_{r-n+1}-\widetilde{\Psi}_{r-n})(\widetilde{\Psi}_{m-r}-\widetilde{\Psi}_{m-r-1}) 
\label{4}
\end{gather}
which is precisely Eq. (\ref{e2}) with the replacement of $f$ by $\widetilde{\mathcal{F}}$. For the defective system made by introducing the rate $f$ drawn from its probability distribution and placing it between the sites $s$ and $s+1$, the propagator is the sum of the propagator given above and an additional term so that the defective propagator is
$$\widetilde{\chi}_{m,n}+\left[\frac{(f-\widetilde{\mathcal{F}})}{1+(f-\widetilde{\mathcal{F}})\widetilde{\beta}}\right](\widetilde{\chi}_{m,s}-\widetilde{\chi}_{m,s+1})(\widetilde{\chi}_{s+1,n}-\widetilde{\chi}_{s,n})
%\label{7}
$$
where, for notational convenience we have introduced the abbreviation $$\widetilde{\beta}=-\widetilde{\chi}_{s+1,s}+\widetilde{\chi}_{s+1,s+1}+\widetilde{\chi}_{s,s}-\widetilde{\chi}_{s,s+1}.$$ The second term in the defective propagator describes the modification by the barrier lying between the sites $s$ and $s+1$. We get a different solution for every ensemble member, the difference being in the value of $f$. We require the self-consistency condition that the ensemble average over the $f$'s give us simply $\widetilde{\chi}_{m,n}$. This must be true whatever the $n$ in the propagator or whichever barrier $s$ characterizes. Therefore, the ensemble average of the factor in the square brackets in the propagator expression above must vanish. This provides a prescription for obtaining the effective quantity $\widetilde{\mathcal{F}}$ through the solution of the implicit equation
\begin{equation}
\int df\rho(f)\left[\frac{f-\widetilde{\mathcal{F}}}{1+(f-\widetilde{\mathcal{F}})\widetilde{\beta}}\right] =0.
\label{mainresult}
\end{equation}
The chain details are reflected in $\widetilde{\beta}$ and the randomness of the $f$'s in $\rho$. There is no $f$-dependence in $\widetilde{\mathcal{F}}$ and $\widetilde{\beta}$, although $\widetilde{\beta}$ is a function of $\epsilon$ as well as of $\widetilde{\mathcal{F}}(\epsilon)$. Because of the independent dependence of $\widetilde{\beta}$ on $\epsilon$, the solution of Eq. (\ref{mainresult}) yields an explicit $\epsilon$-dependence of the effective quantity $\widetilde{\mathcal{F}}(\epsilon)$ that we seek. Different probability distributions result in different expressions for the effective medium quantity $\widetilde{\mathcal{F}}.$

For long times, the Markoffian approximation of the memory function is appropriate. This involves taking the $\epsilon \rightarrow 0$ limit of the Laplace transforms in the expressions above. We first see from Eq. (\ref{4}) that, in this limit, 
$$
\widetilde{\beta}=-\widetilde{\chi}_{s+1,s}+\widetilde{\chi}_{s+1,s+1}+\widetilde{\chi}_{s,s}-\widetilde{\chi}_{s,s+1} \rightarrow \frac{1}{\widetilde{\mathcal{F}}(0)}.
$$

This remarkable simplification allows us to conclude that, in the long time approximation, the effective value of the intercompartmental rate is given trivially as the reciprocal of the ensemble average of the reciprocals of individual intercompartmental rates:
\begin{equation}
\frac{1}{f_{eff}}=\frac{1}{\widetilde{\mathcal{F}}(0)}=\int df\frac{\rho(f)}{f}.
\label{feff}
\end{equation}

Application of effective medium theory has thus reduced the disordered problem of interest in the present paper into the ordered effective problem which was completely analyzed in ref. \cite{KGK}. Combining that analysis with Eq. (\ref{feff}), we find that the overall effective transfer rate for the diffusion of the molecule (taking into account both the existence of the compartments of size $H$ and the existence of disorder in the rates $f$) is given by
\begin{equation}
F_{eff}=\frac{\left(\frac{H+1}{H}\right)}{\left[\frac{1}{F}+\frac{1}{H}\int df\frac{\rho(f)}{f}\right]}.
\label{effdiff}
\end{equation}
For large $H$ (for instance if $H>>1$), Eq. (\ref{effdiff}) simply states that the effective overall transfer rate is the harmonic mean of the intracompartment rate $F$ and the effective intercompartmental rate $f_{eff}$ reduced by the size of the compartments. 

\subsection{Specific Cases}
Equation (\ref{effdiff}) is one of our main results for the case in which disorder appears only in the values of the intercompartmental rates: it allows us to translate the randomness of the intercompartmental rates as expressed in the form of the distribution $\rho(f)$ directly into the effective diffusion parameters of the system. We will now consider several different cases of $\rho(f)$ for purposes of illustration. 

Two trivially expected results emerge in a straightforward fashion: If $\rho(f)$ has a non-zero value at $f=0$ so that there is a non-zero probability of having disconnected sites, the effective hopping rate at long times must vanish. This is clear from Eq. (\ref{effdiff}) since $\rho(f)$ can be written as $\rho(f)=\delta(f)+R(f)$ where $R(f)$ is some distribution of $f$ obeyed for all $f$ except for $f=0$. The integral in the denominator of Eq. \ref{effdiff} diverges, giving $F_{eff}=0$. Similarly, if there is only a single value of the intercompartmental rate $f$, viz. $g$: $\rho(f)=\delta(f-g)$, Eq. (\ref{effdiff}) reduces to the corresponding expression in the analysis of the \emph{ordered} system treated in ref. \cite{KGK}.

For the case when there are two values of the intercompartmental rate appearing with different weights: $$\rho(f)=A_1\delta(f-f_{1})+A_2\delta(f-f_{2}),$$ where obviously the normalization is $A_1+A_2=1,$ the overall effective rate  is given by
%\begin{equation}
%F_{eff}=f_{1}f_{2}\frac{H+1}{(f_{1}+f_{2})/2+H(f_{1}f_{2})/F}
%\label{17}
%\end{equation}
\begin{equation}
F_{eff}=\frac{H+1}{\frac{H}{F}+\frac{A_2f_1+A_1f_2}{f_1f_2}}.
\label{twofs}
\end{equation}
Note that, if either $f_{1}$ or $f_{2}$ is zero, $F_{eff}$ vanishes as we have stated above. 

Finally, we will exhibit the continuum limit of our results, selecting three specific distributions for the intercompartmental rates. We display the continuum limit expressions because, particularly for the problem of molecular motion in cell membranes, they are more directly applicable than their more general discrete counterparts. The  continuum limit means that the lattice constant $a \rightarrow 0,$ transforming hops among discrete sites on a chain to flow on a continuous line. As is well-known, and explicitly commented on elsewhere, (see, e.g., refs. \cite{KGK,simpson}), as $a \rightarrow 0$, it is necessary that  $f,H\rightarrow \infty$ as $1/a$ but $F\rightarrow \infty$ as $1/a^2.$ With this appropriate limiting behavior, the overall effective diffusion constant is given as
\begin{gather}
\lim_{a \rightarrow 0}F_{eff}a^{2}=D_{eff}=\frac{D}{1+\frac{D}{L}\int d\mathscr{D}_{f}\frac{\rho\left(\mathscr{D}_{f} \right)}{\mathscr{D}_{f}}} \label{limit} \\
\mathscr{D}_{f}=\lim_{a \rightarrow 0}fa \nonumber
\end{gather}
where $D$ is the continuum limit of $Fa^{2}$, $\mathscr{D}_f$ is the continuum limit of $fa$, and $L,$ the continuum limit of $(H+1)a,$ is the size of the compartment.  Factors such as $(H+1)/H$ collapse into $1.$ The ratio $\mathscr{P}=\mathscr{D}_f/D$ is what is sometimes called \cite{powlesetal} the permeability.

To illustrate the effect of the form of the distribution functions, we now consider several explicit realizations of $\rho(\mathscr{D}_f)$. We evaluate Eq. (\ref{limit})  for the three respective cases of a constant distribution in an interval, a biased distribution that peaks at a value related to the spread of the distribution, and a biased distribution that peaks at a value independent of the distribution spread. We use the normalization $\int \rho\left(\mathscr{D}_{f} \right)d\mathscr{D}_{f}=1.$

\subsubsection{Uniform distribution}
If $\rho\left(\mathscr{D}_{f} \right)$ is a non-zero constant in an interval of values of $\mathscr{D}_{f}$, i.e., for $l<\mathscr{D}_{f}<u$, and vanishes otherwise, then we have from Eq. (\ref{limit})
\begin{gather}
%D_{eff}=\frac{L(u-l)}{\frac{L(u-l)}{D}+\ln\left( u/l \right)} .
\frac{D_{eff}}{D}=\left[1+\frac{D\ln(u/l)}{L(u-l)}\right]^{-1}.
\end{gather}

\subsubsection{Rayleigh distribution}
If $\rho\left(\mathscr{D}_{f} \right)$ is a biased Gaussian, called sometimes a Rayleigh distribution:
\begin{equation}
\rho(\mathscr{D}_f)=\frac{\mathscr{D}_{f}e^{-\mathscr{D}_{f}^{2}/2\sigma^{2}}}{\sigma^{2}} ,
\end{equation}
where the mean is $\sigma\sqrt{\pi/2}$, and the variance $\sigma^{2}\left(\frac{4-\pi}{2}\right)$ is proportional to the square of the mean, we have
\begin{gather}
%D_{eff}=\frac{L \sigma}{\frac{L \sigma}{D}+\sqrt{\frac{\pi}{2}}} .
\frac{D_{eff}}{D}=\left[1+\frac{D\sqrt{\pi/2}}{L\sigma}\right]^{-1}.
\end{gather}

\subsubsection{Rice distribution}
To have two independently controllable parameters, one deciding the value at which the distribution peaks, and the other the spread, we consider what is called the Rice distribution:\begin{equation}
\rho(\mathscr{D}_f)=\frac{\mathscr{D}_f}{\sigma^{2}}e^{-\frac{(\mathscr{D}_f^{2}+v^{2})}{2\sigma^{2}}}I_{0}\left(\mathscr{D}_f\frac{v}{\sigma^{2}} \right)
\end{equation}
the two parameters being $\sigma$ and $v.$ The mean is $\sigma\sqrt{\pi/2}L_{1/2}\left(-{v^{2}}/{2\sigma^{2}}\right),$ and the variance is $2\sigma^{2}+v^{2}-{\pi\sigma^{2}}/{2}L_{1/2}^{2}\left(-{v^{2}}/{2\sigma^{2}}\right).$ Here, $L_{1/2}(x)=e^{x/2}\left[ (1-x)I_{0}(-x/2)-xI_{1}(-x/2) \right]$ is the Laguerre Polynomial of fractional order and $I_m(x)$ are modified Bessel Functions of the first kind. We find
%the two parameters being $\sigma$ and $v.$ The mean is $\sigma\sqrt{\pi/2}L_{1/2}\left(-\frac{v^{2}}{2\sigma^{2}}\right),$ and the variance is $2\sigma^{2}+v^{2}-\frac{\pi\sigma^{2}}{2}L_{1/2}^{2}\left(-\frac{v^{2}}{2\sigma^{2}}\right).$ Here, $L_{1/2}(x)=e^{x/2}\left[ (1-x)I_{0}(-x/2)-xI_{1}(-x/2) \right]$ is the Laguerre Polynomial of fractional order and $I_m(x)$ are modified Bessel Functions of the first kind. Then we have
\begin{gather}
%D_{eff}=\frac{L \sigma}{\frac{L \sigma}{D}+ \sqrt{\frac{\pi}{2}}e^{-v^{2}/4 \sigma^{2}}I_{0}( v^{2}/4 \sigma^{2})} .
\frac{D_{eff}}{D}=\left[1+\frac{D\sqrt{\pi/2}}{L\sigma}e^{-v^2/4\sigma^2}I_0(v^2/4\sigma^2)\right]^{-1}.
\end{gather}
The distributions are plotted in Fig. 2. Note that the Rice distribution appears highly symmetric around its peak value although, like the Rayleigh distribution, it has the value $0$ at the value $\mathscr{D}_f=0.$

Notice that, in every case, the expression for the effective diffusion constant depends in essentially the same manner on the ratio of the system (barrier-less) diffusion constant $D$ to the product of the compartment length and a characteristic $\mathscr{D}_f$ value. Except for numerical factors, the latter product is $(u-l)/\ln(u/l)$ for the uniform distribution, $\sigma/\sqrt{\pi/2}$ for the Rayleigh distribution, and $\sigma/\left[\sqrt{\pi/2}e^{-v^2/4\sigma^2}I_0(v^2/4\sigma^2)\right]$ for the Rice distribution.
\begin{figure}[H]
\centering
\includegraphics[width = 8.0cm]{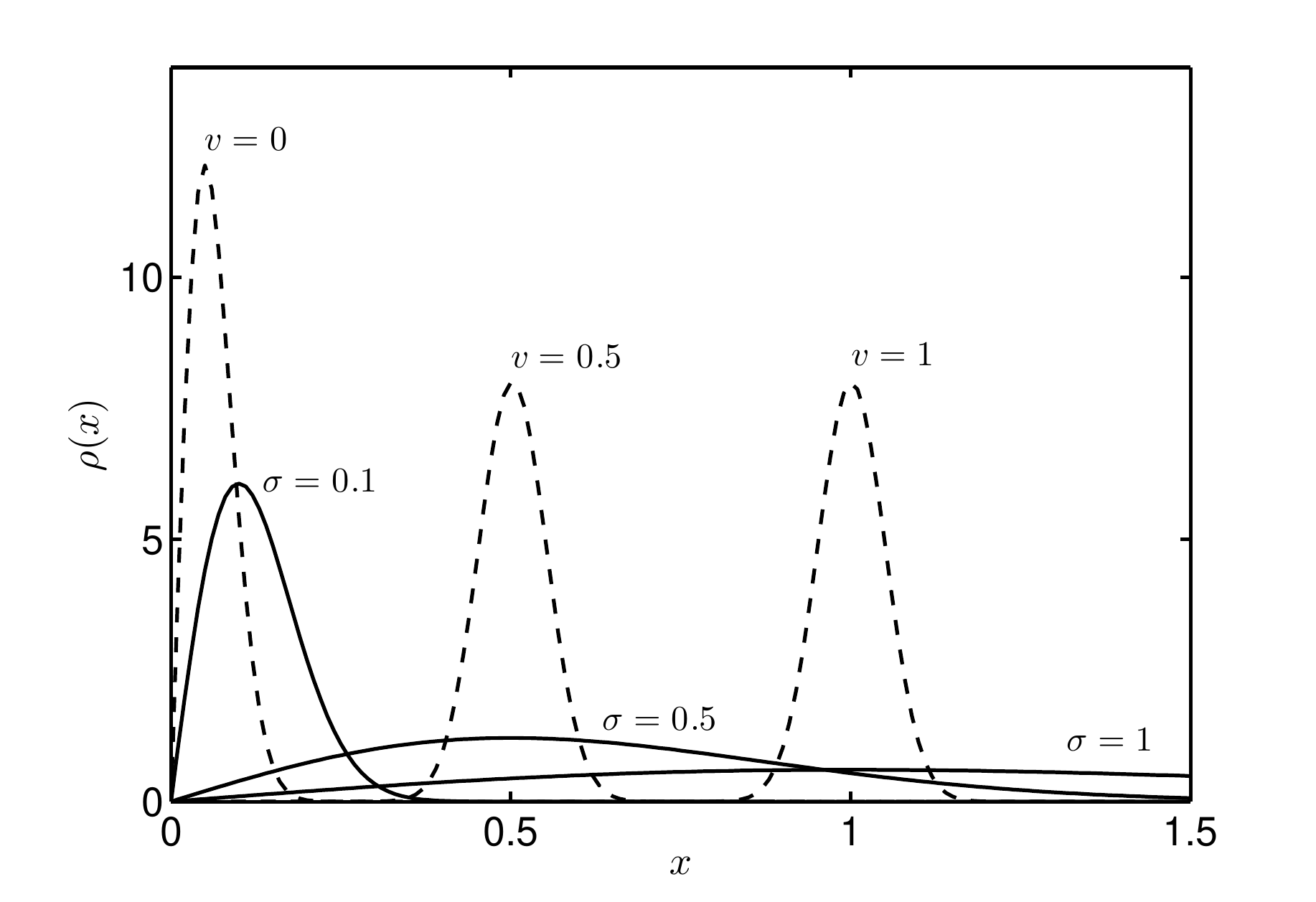}
\caption{Plots of Rayleigh(solid) and Rice(dashed) distributions for various parameter values. For the Rice distribution, $\sigma$=0.05 in all cases. Distributions are normalized such that $\int_{0}^{\infty}dx \rho(x)=1.$ Here, $x$ represents $\mathscr{D}_f.$}
\label{figDistributions}
\end{figure}

\subsection{Determination of the Memory Function}
\label{DetMem}
The full exploitation of the convolution in Eq. (\ref{teff}) and of the consequent memory effects in the motion has seldom been carried out in the literature. The few exceptions are in the context of percolative systems in which the long time diffusion constant vanishes at the percolation point \cite{odagakilax}, of the power law tail analysis given by one of the present authors \cite{parris}, and prescriptions provided by another of the present authors \cite{kenkre} for the determination of memory functions for stress distribution in granular compacts.  In this subsection we briefly show how to explicitly calculate the memory function $\mathcal{F}$ in the Laplace domain following a  prescription similar to that given in the last of the above references \cite{kenkre}. Equations (9) and (10) of ref. \cite{kenkre} should be compared to (\ref{mainresult}) above and (\ref{feff}) below in the present analysis. We recall that the propagators $\Psi_l(t)$ for the original problem (nearest neighbor rates $F$ and no barriers) are $I_l(2Ft)e^{-2Ft}$ and therefore their Laplace transforms can be written in terms of hyperbolic functions as given in Eq. (\ref{hyperbolic}). This allows us to write
%$$\widetilde{\Psi}_{0}-\widetilde{\Psi}_{1}=$$
%$$\sum_{r}\widetilde{\Psi}_{s-r}^{2}=\frac{\coth \xi(H+1)}{4F^{2}\sinh^{2}\xi}$$
%(is this summation over r primed or not?)
%and therefore to 
%\begin{equation}
%\frac{ac}{ad-bc}=\int df\frac{\rho(f)}{f+d/c} 
%\label{12}
%\end{equation}
%where
%\begin{gather}
%a=-\widetilde{\mathcal{F}}\widetilde{\mu}+(1+\widetilde{\mu}F)\nonumber \\
%b=\widetilde{\mathcal{F}}^{2}\widetilde{\mu}-\widetilde{\mathcal{F}}(1+\widetilde{\mu}F) \nonumber \\
%c=\widetilde{\mathcal{F}}\theta\widetilde{\mu}-\zeta-\theta(1+\widetilde{\mu}F) \nonumber \\
%d=-\widetilde{\mathcal{F}}^{2}\theta\widetilde{\mu}+\widetilde{\mathcal{F}}(\zeta-\widetilde{\mu}+\theta(1+\widetilde{\mu}F))+(1+\widetilde{\mu}F) \nonumber 
%\end{gather}
%In order to write the expressions in a compact way, we have defined:
%\begin{gather}
%\theta=\frac{Y-1}{F}-\zeta \nonumber \\
%\zeta=\frac{1}{\widetilde{\mu}F^{2}}(1-2Y+Y^{2}\coth \xi(H+1)) \nonumber \\
%Y=\frac{\cosh\xi - 1}{\sinh\xi} 
%\end{gather}
\begin{figure}[H]
%\centering
\includegraphics[width = 8.0cm]{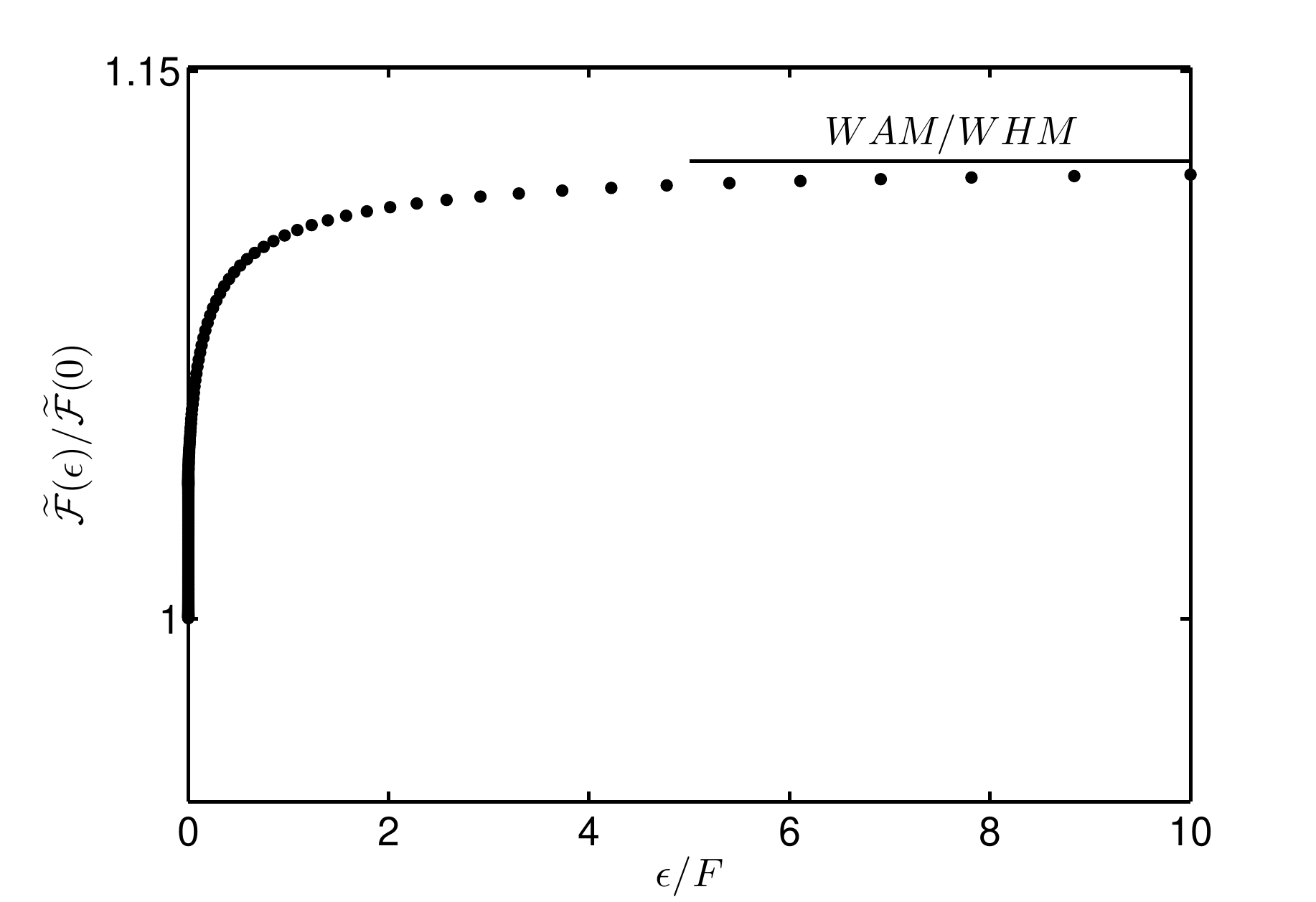}
\caption{Evaluation of the memory function produced by our effective medium theory. Plotted is the $\epsilon-$dependence of the normalized $\widetilde{\mathcal{F}},$ taking $\epsilon$ real for simplicity in display and normalizing it to $F$. Shown is  the case when the distribution of intercompartmental rates $f$ is a sum of two weighted delta-functions: the rates are either $f_1$ or $f_2$. Our evaluation shows that $\widetilde{\mathcal{F}}$ equals the (weighted) arithmetic mean (WAM) of the two rates for large $\epsilon$ and their (weighted) harmonic mean (WHM) for small $\epsilon$, the latter representing the effective long time intercompartmental rate, as expected. The parameter values chosen for this plot are: $f_1=0.1F$, $f_2=0.2F$, $H=10$, $\alpha=0.5$.}
\label{memoryeval}
\end{figure}
\begin{gather}
\widetilde{\beta}=2(\widetilde{\Psi}_{1}-\widetilde{\Psi}_{0})  \\ -\frac{g-\widetilde{\mathcal{F}}}{F^{2}(1+(g-\widetilde{\mathcal{F}})\widetilde{\mu})}\left( 1-2\epsilon\widetilde{\Psi}_{0}+\epsilon^{2}\sum_{r}\widetilde{\Psi}_{s-r}^{2} \right) , \nonumber \\
\sum_{r}^{\prime}\widetilde{\Psi}_{s-r}^{2}=\frac{\coth \xi(H+1)}{4F^{2}\sinh^{2}\xi} . \\
\label{11}
\end{gather}
In the light of this expression for $\widetilde{\beta}$, Eq. (\ref{mainresult}) yields:
\begin{equation}
\frac{1}{\Gamma}=\int df\frac{\rho(f)}{f + (\Gamma - \widetilde{\mathcal{F}})} ,
\label{12}
\end{equation}
where
\begin{gather}
\Gamma=-\frac{(1+\widetilde{\mu}F)-\widetilde{\mathcal{F}}\widetilde{\mu}}{\zeta+\theta( (1+\widetilde{\mu}F) -\widetilde{\mathcal{F}}\widetilde{\mu} )} .
\end{gather}
In order to write the expressions in a compact way, we have defined:
\begin{gather}
\theta=\frac{\coth(\xi/2)-1}{F}-\zeta , \nonumber \\
\zeta=\frac{1}{\widetilde{\mu}F^{2}}(1-2\coth(\xi/2)+\coth^{2}(\xi/2)\coth \xi(H+1)) . \nonumber \\
\end{gather}

We can now solve for $\widetilde{\mathcal{F}}$ for a given $\rho(f)$ by using Eq. (\ref{12}) as outlined for a different case in ref. \cite{kenkre}. As a special case of the distribution we use the case when the intercompartmental rate takes on one of two values with different weights: 
\begin{equation}
\rho(f)=\alpha\delta(f-f_{1})+(1-\alpha)\delta(f-f_{2}).
\label{DoubleDelta}
\end{equation}
Then Eq. (\ref{12}) gives us, for $\widetilde{\mathcal{F}},$
\begin{equation}
\widetilde{\mathcal{F}}^{3}+b \widetilde{\mathcal{F}}^{2}+c \widetilde{\mathcal{F}}+d =0 , 
\label{13}
\end{equation}
where
\begin{align}
b &=  - (f_1 + f_2 - 1/\theta) -  \eta/\theta \widetilde{\mu}, \nonumber \\
c &= f_1 f_2 - [ \widetilde{\mu} (f_2 + \alpha (f_1 - f_2)) + (1+\widetilde{\mu}F) \nonumber \\&- \eta (f_1 +f_2) ]/ \theta \widetilde{\mu} \nonumber, \\
d &=[  \eta f_1 f_2 -  (f_2 + \alpha (f_1 - f_2)) (1+\widetilde{\mu}F) ]/ \theta \widetilde{\mu}, \nonumber \\
\eta & =  (\zeta + \theta(1+\widetilde{\mu}F) ).
\end{align}

The numerical solution of the cubic equation yields the memory function explicitly in the Laplace domain. It is plotted in Fig. 3. We see that, at $\epsilon=0$, $\widetilde{\mathcal{F}}$ tends to the value  of its Markoffian approximation which is the weighted harmonic mean of the two rates $f_1$ and $f_2$. We also see that it tends to the weighted arithmetic mean of the two rates as $\epsilon\rightarrow \infty.$ The intermediate behavior corresponds to intermediate times. We defer to a forthcoming publication \cite{kkp} a detailed application of these and related memory developments.

\section{Disorder in Barrier Placement and in intercompartmental Rates}

We now turn our attention to a generalization of the problem which involves disorder in the placement of the barriers rather than in their heights. We first take the latter to be the same throughout the chain.

\subsection{Disorder in Barrier Placement Only}
\label{disorderplacement}
Let us consider the following distribution for barrier heights: 
\begin{equation}
\rho(f)=\alpha \delta(f-g)+(1-\alpha)\delta(f-F) .
\label{dichotomous}
\end{equation}
Here, the intercompartmental transition rate is either $g$ or $F$ with probabilities $\alpha$ and $1-\alpha$ respectively. When the barrier heights are distributed according to Eq. (\ref{dichotomous}), it is as if, starting with exactly the same system as in ref. \cite{KGK} (periodic barriers of the same height $g$), we replace the barriers randomly and independently with probability $1-\alpha$ with links that have transfer rates $F$, which is the intracompartmental transfer rate. Therefore, the barriers between some of the compartments are now removed which means that some of the compartments are merged. The compartment sizes always are  in multiples of $(H+1)a$ (the smallest compartment size) where $a$ is the lattice constant. In this system, the compartment size, which is the distance between the consecutive barriers, will be a random variable whose distribution will depend on the parameter $\alpha$.

The distribution of distance between two consecutive barriers is easy to calculate. Consider $N$ points in a discreet linear space. To each point we will assign a number $s_{i}$ which is either 1 or 0. We will think of the points in this space as the intercompartmental links in our original problem. In this picture, a point $i$ with $s_{i}=1$ will represent a link with transfer rate $F$ and a point with $s_{i}=0$ will stand for a link with transfer rate $g$, which we have been calling a barrier. Then according to Eq. (\ref{dichotomous}), 0's will occur with probability $\alpha$ and 1's with $1-\alpha$. If we let $\sigma$ be the number of elements in a contiguous sequence of 1's, the distance between two consecutive barriers is simply given by $(\sigma+1)(H+1)a$. Note that $\sigma=0$ corresponds to the case in which the distance between two consecutive barriers is the maximum value $(H+1)a$. When all these arguments are taken into account, the number distribution of $\sigma$ is found to be given by

\begin{gather}
\mathscr{N}(\sigma)=\delta_{\sigma,0}\sum_{j=1}^{N-1}(1-s_{j})(1-s_{j+1}) \nonumber \\
+(1-\delta_{\sigma,0})\sum_{j=1}^{N-\sigma-1}(1-s_{j})\left(\prod_{i=0}^{\sigma-1}s_{j+i+1} \right)(1-s_{j+\sigma+1}) , 
\label{numberdistr}
\end{gather}
where $\mathscr{N}(\sigma)$ is the number distribution of $\sigma$ in a \textit{particular realization} of a 1-D chain as described in the beginning of this section. The first and second terms in Eq. (\ref{numberdistr}) count the occurrence of compartments of sizes $(H+1)a$ and $(\sigma+1)(H+1)a$ respectively. As $s_{i}$'s are independently distributed, we can write:
\begin{equation}
\langle{s}_{i}\rangle =1-\alpha ,
\end{equation}
where the angular brackets mean an ensemble average over all realizations of 1-D chains with intercompartmental transition rates sampled from Eq. (\ref{dichotomous}). Then the ensemble averaged number distribution is given by

\begin{gather}
\langle\mathscr{N}(\sigma) \rangle=(N-\sigma-1)\alpha^{2} (1-\alpha)^{\sigma} ,
\end{gather}
and therefore the probability distribution for $\sigma$ is
\begin{equation}
\langle P(\sigma) \rangle=\frac{\langle \mathscr{N}(\sigma)\rangle}{\sum_{\sigma=0}^{N-1}\langle \mathscr{N}(\sigma)\rangle} .
\end{equation}
As we are interested in particularly in infinite chains, we take the limit $N\rightarrow \infty$. The probability distribution for $\sigma$ becomes
\begin{gather}
\langle P_{N\rightarrow \infty}(\sigma)\rangle= \alpha (1-\alpha)^{\sigma} .
\label{sigmadistr}
\end{gather}
Then the ensemble averaged compartment size distribution would be given by
\begin{equation}
P(\sigma,\alpha)= \alpha (1-\alpha)^{\frac{q}{H+1}-1},
\label{PC}
\end{equation}
where $q=(\sigma+1)(H+1)$ is the dimensionless compartment size. The mean and variance of $P(\sigma,\alpha)$ are given by
\begin{gather}
\overline{q}=\frac{H+1}{\alpha} , \\
(\Delta q)^{2}=\overline{(q^{2})}-\overline{(q)}^{2}=(H+1)^{2}\frac{1-\alpha}{\alpha^{2}} . 
\end{gather}
\begin{figure}
	\centering
		\includegraphics[width=8cm]{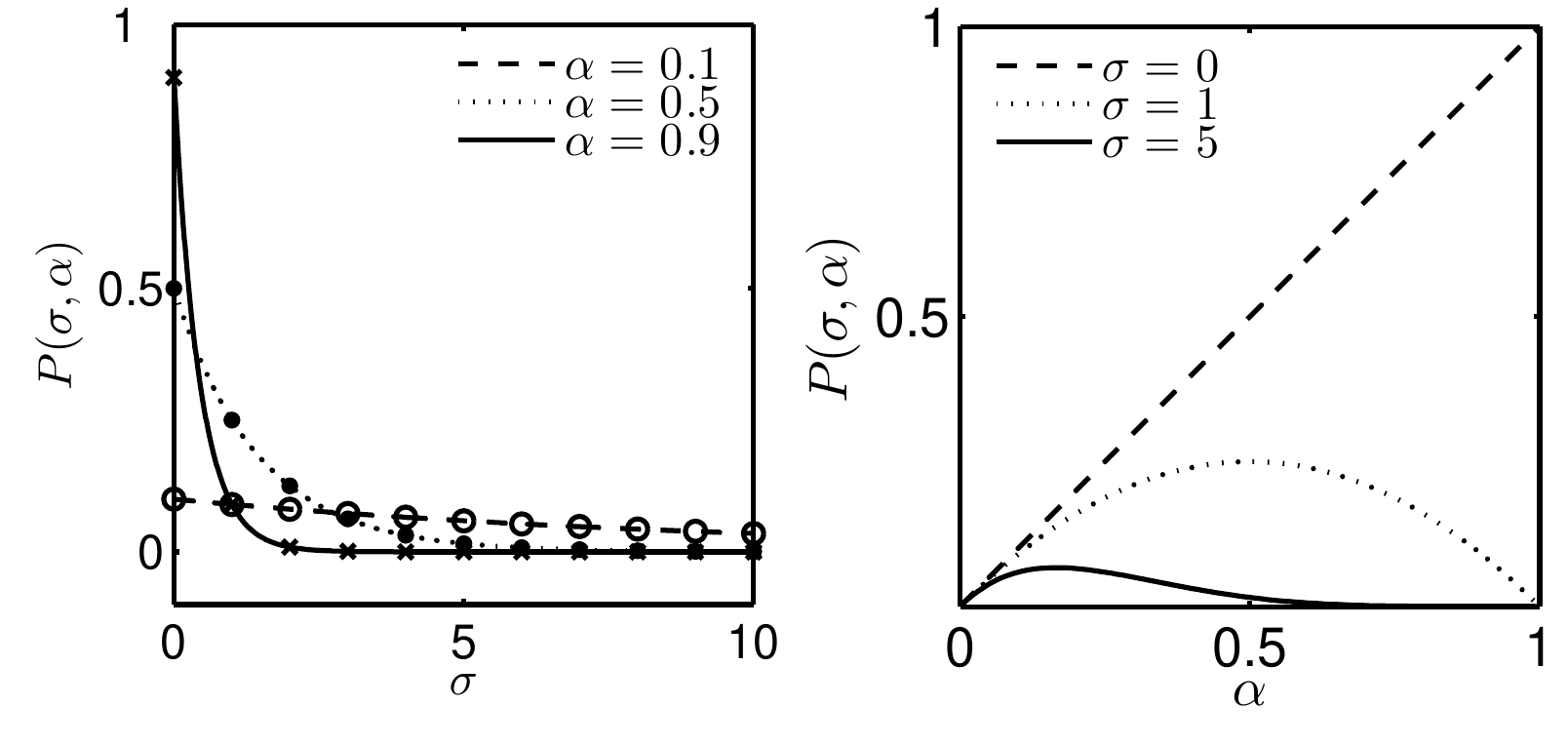}
		\caption{The ensemble averaged probability distribution $P(\sigma,\alpha)$, in the limit as $N\rightarrow \infty$, as a function of $\sigma$ (left) and as a function of its parameter $\alpha$ (right). The continuous lines in the plots correspond to our formula in Eq. (\ref{PC}).}
		\label{Psigma}
\end{figure}

In Fig. \ref{Psigma} we display this distribution: plotted against $\sigma$ for three values of $\alpha$ on the left side and against $\alpha$ for three values of $\sigma$, as shown. The left plot shows that, in all cases the distribution is peaked at vanishing compartment size. The right plot shows that, with the exception of the case $\alpha=0$,  the distribution vanishes for both extremes $\alpha=0$ and $\alpha=1$, but rises and drops for intermediate values.

The general theory we have developed in Section II now provides us, in light of the distribution we have obtained above, with a prescription to calculate the effective long time transfer rate $F_{eff} $ in terms of  the mean (dimensionless) compartment size $\overline{q}$ and the rates $F$ and $f$:
\begin{equation}
F_{eff} = \frac{\overline{q}}{1/f+(\overline{q}-1)/F}.
\label{feff}
\end{equation}

In the continuum limit, obtained by multiplying $F_{eff}$ by $a^2$ and letting $a$ tend to zero appropriately, we note that $H$ also tends to infinity, producing the limit of $\overline{q}a$ as the mean compartment size $Q$ which has dimensions of length. We get
\begin{equation}
\frac{D_{eff}}{D}=\left[1+\frac{D}{Q\mathscr{D}_{f}}\right]^{-1}.
\label{deff}
\end{equation}
%whose mean and variance given by:
%\begin{gather}
%\overline{\sigma}=\frac{1-\alpha}{\alpha}\\
%(\Delta\sigma)^{2}=\overline{(\sigma^{2})}-\overline{(\sigma)}^{2}=\frac{1-\alpha}{\alpha^{2}}
%\end{gather}

\subsection{Disorder also in the Intercompartmental Rates}
\label{disorderboth}
We now give expressions for the effective hopping rate and diffusion constant when both the heights and places of the barriers are random. 

Consider
\begin{equation}
\rho(f)=(1-\alpha) \delta(f-F) + \eta(f,\alpha) ,
\label{randomeverything}
\end{equation}
where $\eta(f,\alpha)$ is a distribution normalized to $\alpha$, with the  understanding that $\eta(0,\alpha)=0$. According to Eq. (\ref{randomeverything}), and the development in Section \ref{disorderplacement}, a fraction $\alpha$ of the intercompartmental links are barriers whose heights are sampled from the distribution $\eta(f,\alpha)$ and the rest are just intracompartmental links with transition rates $F$ that in turn give rise to the variability in compartment sizes. Note that the statistics of different compartment size distributions do not change even if the barrier heights are not the same. Therefore the compartment size distribution can still be obtained from Eq. (\ref{sigmadistr}). Thus we get

\begin{equation}
F_{eff}=\frac{ \frac{H+1}{H} }{ \frac{1}{F}\left( \frac{H+1-\alpha}{H} \right) + \frac{1}{H}\int df\frac{\eta(f,\alpha)}{f} },
\label{randeveryefff}
\end{equation}
and the effective diffusion constant in the continuum limit becomes
\begin{equation}
%D_{eff}=\frac{ 1 }{ \frac{1}{D} + \frac{1}{L}\int d\mathscr{D}_{f}\frac{\eta(\mathscr{D}_{f},\alpha)}{\mathscr{D}_{f} } },
\frac{D_{eff}}{D}=\left[1+\frac{D}{L}\int d\mathscr{D}_{f}\frac{\eta(\mathscr{D}_{f},\alpha)}{\mathscr{D}_{f}}\right]^{-1}.
\label{randeveryeffdiff}
\end{equation}
Here, $\int_{0}^{\infty}dx \eta(x,\alpha)=\alpha$, $x$ being $f$ and $\mathscr{D}_{f}$ in the two respective equations above. Note that when $\alpha=0$, so that there are no barriers, $\eta(x,0)$ vanishes identically as it is a positive function, and the results reduce to $F_{eff}=F$ and $D_{eff}=D$.

\section{Conclusion}
\label{conclusion}
This paper has a twin purpose, an investigation of the effect of disorder on molecular motion in cell membranes, and the development of effective medium theory approaches in new practical directions. We have discussed the first context briefly in the Introduction. Single fluorescent video imaging \cite{sfvi1,sfvi2} and single particle tracking \cite{spt1,spt2} are the two most common ways of observing laterally moving molecules on live cell membranes. The latter can provide information about motion at very short times such as confinement effects. Our theory is positioned to address the results of both measurement techniques. Our results allow us to describe the effect of disorder both in the location and the size of fence barriers, and the motion of both proteins and lipids in cell membranes. These barrier parameter variations arise from the dynamics of the cytoskeleton and of the transmembrane molecules. Observational determination of such variations should be possible through the use of modern methods involving optical tweezers \cite{tweezers} and electron tomography techniques \cite{tomo} that enable the imaging of the actin filaments and the cell surface. These experimental data can be used in conjunction with the calculations we have presented to extract information about compartment sizes and their spatial fluctuations, as well as about the energetic barrier magnitudes and their fluctuations. As more detail becomes available about the membrane dynamics from experiment, it will be straightforward to give further quantitative shape to the theory we have provided here.

The second context of our study lies in presenting effective medium calculations in systems which are partly ordered and partly disordered. This line of research has been recently taken by two of the present authors in their study of transport on small world networks, particularly of the Neumann-Watts kind \cite{parriskenkre,cpk,pck}. In those systems standard rings (finite chains with periodic boundary conditions) with nearest neighbor hopping rates for the random walker form the ordered part and additional small world connections make up the disordered part. Effective medium theories developed for those systems envisage the effective system as the rings with additional \emph{periodically placed} connections. Here, for the cell membrane problem, the standard chain forms the ordered part, and the disordered barriers (in magnitude and location) the disordered part. Our effective medium theory here takes the effective system as the chain with periodically placed barriers of constant magnitude. 

Our starting point in this paper has been Eq. (\ref{tougheqn}) which is impossible to solve in practice because of the enormous number of irregularities. Effective medium considerations have yielded as a general result the prescription (\ref{mainresult}) to calculate $\widetilde{\mathcal{F}}$ appearing in the effective medium Eq. (\ref{teff}) as a function of the Laplace variable $\epsilon,$ starting from known random distributions. We have illustrated the procedure to carry out this prescription for a simple case which results in a cubic equation for $\widetilde{\mathcal{F}}$, viz. (\ref{13}). Its solution is in Fig. \ref{memoryeval} and shows physically expected limiting results. Long time approximations to the memory $\mathcal{F}(t)$ are appropriate if the interest lies in long time description of the transport of the random walker (e.g., of the molecules in the cell membranes.) In such a case we have developed a quite general result, Eq. (\ref{limit}), which transparently connects the distribution of intercompartmental rates $\rho(f)$ to the effective diffusion constant. We have illustrated these results for several specific cases of the distribution $\rho(f)$ including a Rayleigh and a Rice distribution, and also employed the general development to shed light on effects of disorder in barrier placement as well as in intercompartmental rates.

\begin{figure}
	\centering
		\includegraphics[width=8cm]{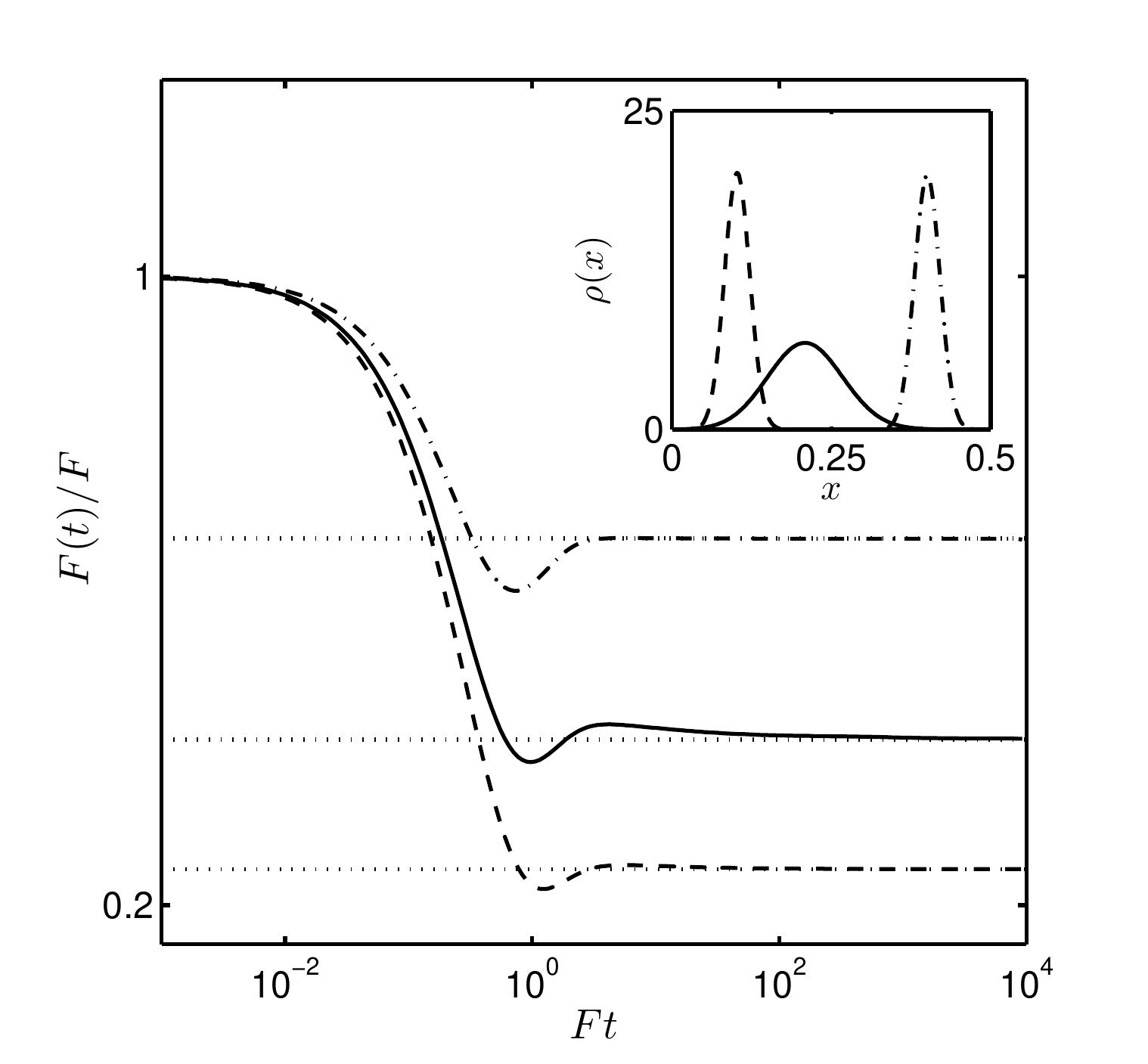}
		\caption{Excellent agreement for sufficiently long times of our effective medium theory with numerically obtained solutions of the Master equation with disordered barrier heights typified by three Rice distributions with different parameters, ($v=0.1$, $s=0.02$), ($v=0.2$, $s=0.06$) and ($v=0.4$, $s=0.02$), respectively denoted by dashed, solid and dash-dotted curves in both the main figure which depicts the instantaneous transfer rate $F(t)$ normalized to $F$, and in the inset which shows the corresponding probability distributions $\rho$. Long time predictions (effective transfer rates) of our effective medium theory, shown by horizontal dotted lines, are reached asymptotically in each case. See text for other details including the observed dips.}
		\label{verif}
\end{figure}

In concluding, we display a comparison of the results of our effective medium theory with numerical solutions of the disordered Master equation for various realizations of $f$-disorder, with $H$ held constant. We see in Fig. \ref{verif} that the agreement is excellent at sufficiently long times. We considered several different kinds of distribution and found the results to be essentially identical. The results displayed in Fig. \ref{verif} correspond to three Rice distributions with different parameters,  ($v=0.1$, $s=0.02$), ($v=0.2$, $s=0.06$) and ($v=0.4$, $s=0.02$), the probability distribution functions being shown in the inset. The same three kinds of curve,  dashed, solid and dash-dotted curves respectively, used to display the distributions are used correspondingly in the main figure to show the results of the  numerical solution of Eq. (\ref{tougheqn}) followed by performing an ensemble average. The specific plots are of the instantaneous transfer rate $F(t)$, normalized to the barrier-less system transfer rate $F$. This $F(t)$ is one half the time derivative of the (dimensionless) mean square displacement, and should not be confused with the memory $\mathcal{F}(t)$. Two features are visible: dips  below the eventual asymptotic values, and the coincidence of the asymptotic values with horizontal dotted lines that represent our effective medium theory. The dips arise from the fact that the random walker is assumed to start initially at the center of one of the compartments: repeated encounters with walls when the effective transfer rate drops are responsible for the dips. A detailed discussion of the dips is given in ref. \cite{KGK}. The asymptotic coincidence of the numerical solutions with the effective medium theory provides graphical validation of the latter.

\acknowledgments
This work was supported in part by the NSF under
grant no. INT-0336343,  by the
Program in Interdisciplinary Biological and Biomedical
Sciences at UNM funded by the Howard Hughes Medical
Institute, and by DARPA under grant no. DARPAN00014-
03-1-0900. One of the authors (VMK) thanks Dr. Luca Giuggioli
for initial discussions on the study of molecular motion on cell membranes.

\end{document}